\documentclass{article}

\usepackage{PRIMEarxiv}

\usepackage[utf8]{inputenc} % allow utf-8 input
\usepackage[T1]{fontenc}    % use 8-bit T1 fonts
\usepackage{hyperref}       % hyperlinks
\usepackage{url}            % simple URL typesetting
\usepackage{booktabs}       % professional-quality tables
\usepackage{amsfonts}       % blackboard math symbols
\usepackage{amsmath}
\usepackage{nicefrac}       % compact symbols for 1/2, etc.
\usepackage{microtype}      % microtypography
\usepackage{lipsum}
\usepackage{fancyhdr}       % header
\usepackage{graphicx}       % graphics
\graphicspath{{media/}}     % organize your images and other figures under media/ folder
\usepackage{ulem}
\usepackage{placeins}
\usepackage{xcolor}
\usepackage{multirow}
\title{Experimental determination of ruthenium L-shell fluorescence yields and Coster-Kronig transition probabilities}

\author{
  Nils Wauschkuhn, Katja Frenzel, Burkhard Beckhoff, Philipp Hönicke \\
  Physikalisch-Technische Bundesanstalt \\
  Abbestr. 2-12 \\
  10587 Berlin\\
  Germany\\
  \texttt{nils.wauschkuhn@ptb.de} \\
}

\begin{document}

\maketitle

\begin{abstract}
The L-shell fluorescence yields and the Coster-Kronig factors of ruthenium (and the corresponding uncertainty) were determined for the first time experimentally by applying radiometrically calibrated instrumentation of the Physikalisch-Technische Bundesanstalt. The resulting fluorescence yields ($\omega_{L_3}=0.0459(20)$, $\omega_{L_2}=0.0415(26)$, $\omega_{L_1}=0.0109(9)$) and the Coster-Kronig factors ($f_{23}=0.177(32)$, $f_{13}=0.528(90)$, $f_{12}=0.173(73)$) agree reasonable well with parts of the data from the literature.
\end{abstract}

% keywords can be removed

\section{Introduction} 
Ruthenium is a versatile and widely used chemical element playing a crucial role in important areas of science and technology. Several applications in the area of semiconductor fabrication or catalysis can be identified, where ruthenium is essential. For extreme ultraviolet lithography masks \cite{WU2021100089, MaskAbsorberEUVL} or as interconnect metal \cite{RutheniumMetallizationInterconnects, EpitaxialBeyondCu, PassivationInterconnects} either ruthenium or ruthenium-containing materials are very relevant. In catalysis, ruthenium-based catalysts provide remarkable properties in several different applications \cite{GRAMAGEDORIA2021213602}. In addition to this, ruthenium is also of relevance for emerging applications for energy storage \cite{Battery, LiBattery} and medicine \cite{Cancer, allardyce2001ruthenium}.

But, if X-ray fluorescence (XRF) based techniques are to be used for determining the ruthenium content in such materials, one quickly finds that the knowledge of the relevant atomic fundamental parameter (FP) data for ruthenium is very limited: For ruthenium, especially its L-shell FP data and namely the L-subshell fluorescence yields and Coster-Kronig factors (CK), no experimentally determined data seems to exist so far. Available data in the literature is either purely theoretically determined or perhaps even less favorable, only interpolated employing adjacent chemical elements. As these FPs quantitatively describe the process of X-ray fluorescence generation, they are very crucial for most quantification approaches in XRF. Thus, they have a direct influence on the accuracy of the XRF quantification results.

As this is a highly inadequate situation, we applied the PTB's reference-free X-ray spectrometry toolset in order to experimentally determine the fluorescence yields and the Coster-Kronig factors of the $L$-subshells of ruthenium for the first time. Based on transmission and fluorescence experiments on thin film samples, such FP data can be derived as already demonstrated for a wide range of chemical elements \cite{M.Kolbe2012, M.Kolbe2015, Kayser_2022, Hoenicke2022, Ta-Paper}.

\section{Experimental procedure}
\label{sec:exp}
For an experimental determination of L-shell fluorescence yields and Coster-Kronig transition probabilities, both fluorescence- and transmission experiments with a selective excitation of the three L-subshells on either a free standing thin foil or a thin coating on a carrier foil are required \cite{M.Kolbe2012, M.Kolbe2015, Kayser_2022, Ta-Paper}. In the present work, these experiments were conducted on the four-crystal monochromator (FCM) beamline \cite{Krumrey1998} of BESSY II using a vacuum chamber that is in-house developed \cite{M.Kolbe2005a}. This chamber was endowed with a silicon drift detector (SDD) of which the detection efficiency is radiometrically calibrated and the response functions are determined experimentally \cite{F.Scholze2009}.
The employed sample was a highly homogeneous 150 nm ruthenium deposition on a 500 nm Si$_3$N$_4$ membrane. To be able to isolate the ruthenium contribution from the total sample transmission, also a blank membrane of nominally identical thickness was used. Any potential moderate variation in the Si$_3$N$_4$ membrane thickness is only a second-order contribution to the uncertainties. Both samples were positioned in the chamber's center by using an x-y-scanning stage.
The angle between the incoming beam and the sample as well as the angle between sample surface and detector was set to 45°.\\

The transmission measurements were conducted in an energy range around the Ru-L absorption edges between 2.1~keV and 4~keV. Furthermore, X-ray fluorescence measurements were performed in the incident-energy domain between 2.8~keV and 3.4~keV.
The established methodolgy \cite{M.Kolbe2012, P.Hoenicke2014, M.Kolbe2015,Menesguen2018} to derive the relevant L-shell FPs from this exerimental dataset is described in the following.

According to the Sherman equation \cite{Sherman1955}, the measured count rate of fluorescence photons of a one-elemental foil, which is irradiated under 45°, is the product of the fluorescence production cross section $\sigma_{Li}$ of the considered shell, the incoming $\Phi_0(E_0)$ as well as the fluorescence photon flux $\Phi^d_i(E_0)$, the detection efficiency of the SDD, the mass deposition of that element, the attenuation correction factor $M_{i,E_0}$ and the solid angle $\Omega$ of detection of the SDD. 

The self-attenuation correction factor takes into account the attenuation %\textcolor{blue}{absorption = tau, mu ist aber tau + scattering} 
of the incident radiation and of the fluorescence radiation on its way through the sample. The corresponding sample-specific attenuation correction factor $M_{i,E_0}$ is determined by transmission experiments taking advantage of the fact, that the knowledge of the ruthenium deposition thickness $d$ and its density $\rho$ is not needed since they appear only in a product with the mass absorption coefficient $\mu_S$ or with the subshell photoionization cross section $\tau_S$. The product $\mu_S \rho d$ is derived from the transmittance data using the Lambert-Beer law. 

For incident energies $E_0$ between the $L_3$ edge and the $L_2$ edge, the fluorescence production factor for the $L_3$-subshell is
\begin{equation}
\sigma_{L3}(E_0)\rho d = \omega_{L3} \tau_{L3}(E_0)\rho d = \frac{\Phi^d_i(E_0)M_{i,E_0}}{\Phi_0(E_0)\frac{\Omega}{4\pi}},
\label{eq:prodCS}
\end{equation}
where $\omega_{L3}$ is the ruthenium L$_3$ fluorescence yield which should be determined. The sample-specific attenuation correction factor $M_{i,E_0}$ is defined as
\begin{equation}
M_{i,E_0} = \frac{(\frac{\mu_S(E_0)\rho d}{\sin \theta_{in}}+\frac{\mu_S(E_i)\rho d}{\sin \theta_{out}})}{(1-\exp[-(\frac{\mu_S(E_0)\rho d}{\sin \theta_{in}}+\frac{\mu_S(E_i)\rho d}{\sin \theta_{out}})])}.
\label{eq:M}
\end{equation}
Here, $\theta_{in}$ is the angle between the incident beam and the sample surface, $\theta_{out}$ is the angle between the sample surface and the SDD detctor.\\

Due to the so-called Coster-Kronig effect, the effective photoionization cross section $\tau_{\mathrm{eff},L i}(E_0)$ for L$_3$ and L$_2$ is a linear combination with the higher bound shells since for photon energies above the excitation energy of the next subshell, created holes in $L_2$ can decay into $L_3$ by ejecting outer electrons. As a result, more than the directly created holes in $L_3$ exist. The CK-factor $f_{23}$ provides the probability for this to happen and similar transitions can occur between the $L_1$ and the $L_2$ and $L_3$ shells. So for an incident photon energy above the $L_1$ threshhold, the fluorescence production factors are defined as: \\

\begin{align}
\tau_{\mathrm{eff},L_3}(E_0) &= \tau_{L3}(E_0) + f_{23}\tau_{L2}(E_0) + [f_{13} +  f_{12} f_{23}] \tau_{L1}(E_0)
\label{eq:Taueff3}\\
\tau_{\mathrm{eff},L_2}(E_0) &= \tau_{L2}(E_0) + f_{12}\tau_{L1}(E_0)
\label{eq:Taueff2}\\
\tau_{\mathrm{eff},L_1}(E_0) &= \tau_{L1}(E_0)
\label{eq:Taueff1}
\end{align}

Here, the $\tau_{L_i}(E_0)$ are the photoionization cross sections of the respective $L_i$ subshell\cite{P.Hoenicke2014}, and $f_{ij}$ are the Coster-Kronig factors. For incident energies below the subsequent subshell, the corresponding subshell photoionization cross section is zero ($\tau_{Li}(E_0)=0$ for $E_0<E_{Li}$). Therefore, the fluorescence yields are determined for energies $E_0$ above the excitation energy of the considered and below the subsequent subshell.

All relevant observables are accessible from the experimental data as $\mu_S \rho d$ are determined for the relevant energies by measuring the transmission of the ruthenium coating. $\Phi^d_i(E_0)$ is determined by spectral deconvolution of the recorded SDD spectra considering the relevant fluorescence lines and relevant background contributions such as bremsstrahlung. $\Phi_0(E_0)$ and $\Omega$ are known because of PTB's calibrated instrumentation \cite{Beckhoff2008}.

\begin{figure}
  \centering
    \includegraphics[width=9cm]{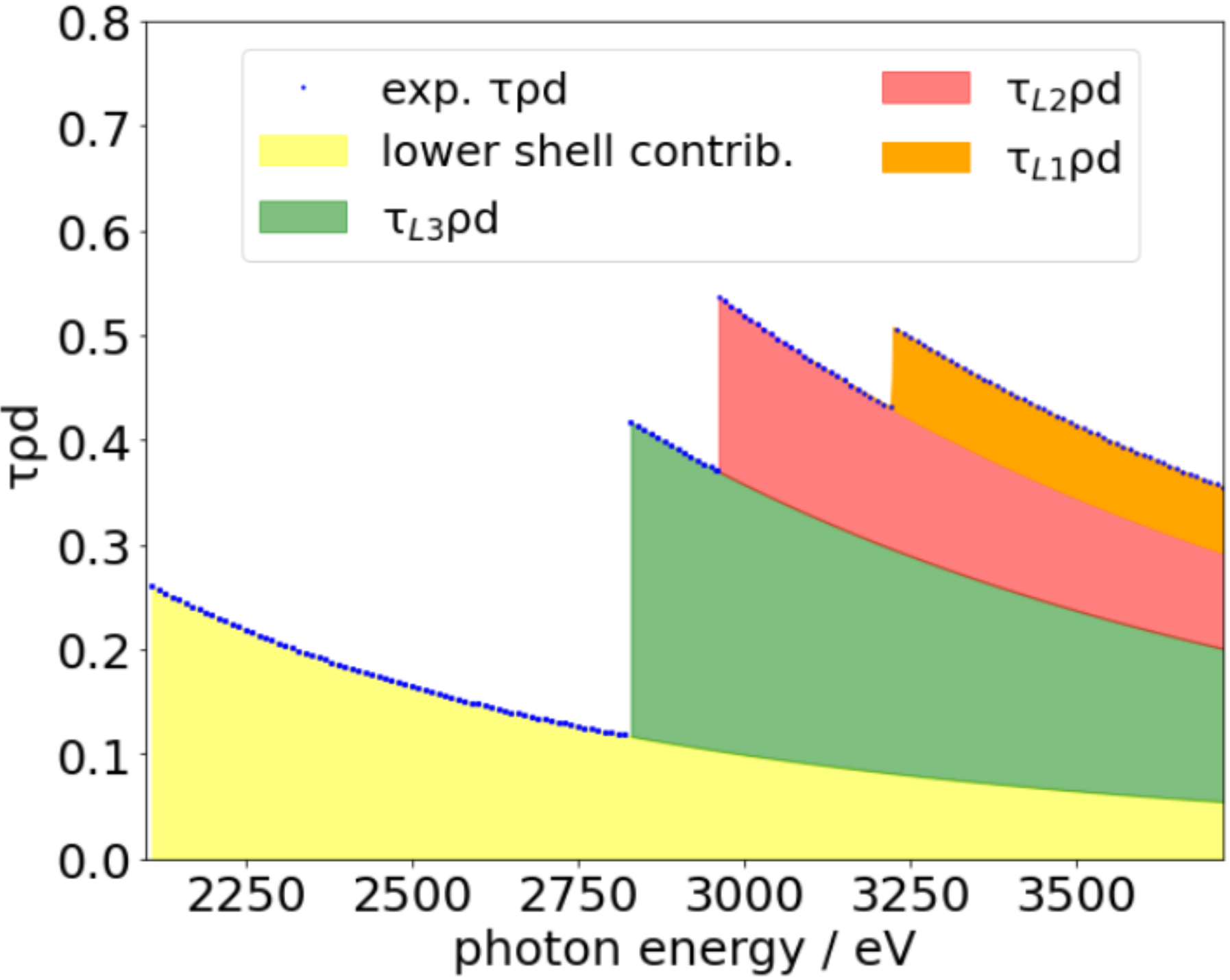}
  \caption{$\tau_S(E_0) \rho d$ determined for the ruthenium thin film: separation of the contributions of lower bound shells (yellow), $L_3$ (green), $L_2$ (red) and $L_1$ (orange)}
  \label{fig:Taurhod}
\end{figure}

Taking into account the theoretical ratio of scattering to ionization cross sections, which one can take from databases\cite{H.Ebel2003}, the sample-specific total photoionization cross section $\tau_{S} \rho d$ can be derived. To isolate the subshell contributions of the different $\tau_{L i}$, Ebel polynomials \cite{H.Ebel2003} for each $L_i$ contribution as well as a total cross section for lower bound shells are scaled into the data (see figure \ref{fig:Taurhod}). For this scaling process, only the datapoints slightly above each absorption edge are used to minimize the effect of the fine structure. 

\begin{figure}
  \centering
    \includegraphics[width=16cm]{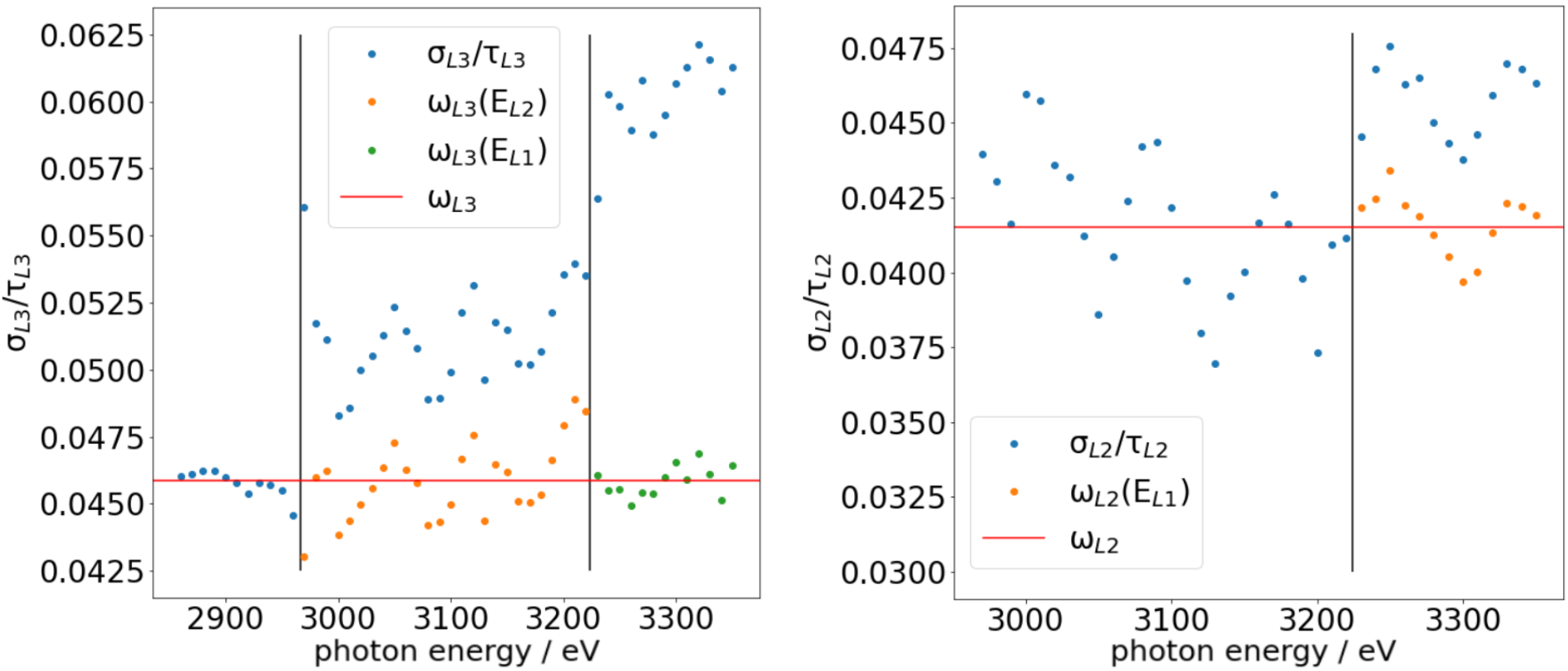}
  \caption{Experimental determination of the Ru-L$_3$ (left image) and Ru-L$_2$ (right image) fluorescence yields: They are determined by averaging over all considered energies where only the respective shell is excited (below $L_2$ for $\omega_{L3}$ and below $L_1$ for $\omega_{L2}$). Using these $\omega_{L3}$ and $\omega_{L2}$, the Coster-Kronig factors are determined in such a way that the average in the higher energy domains matches the $\omega_{Li}$ value.}
  \label{fig:CK-check}
\end{figure}

With these determined $\tau_{L i} \rho d$, the equations for the fluorescence production cross sections can be solved for $\omega_{Li}$.  By replacing $\tau_{L i}$ by the effective photoionization cross section according to equations \ref{eq:Taueff3}-\ref{eq:Taueff1}, eqn \ref{eq:prodCS} can be applied also for energies above the next subshell. Therefore, to determine $f_{23}$, energies between $E_{L3}$ and $E_{L2}$ were considered, see figure \ref{fig:CK-check}: With the already determined $\omega_{L3}$, the modified version of eqn \ref{eq:prodCS} can be solved for $f_{23}$. $f_{12}$ is determined in the same way but applied for the fluorescence of the L$_2$ shell and for $E_{L2} < E_0 < E_{L1}$ with the already determined $\omega_{L2}$. With these determined $f_{23}$ and $f_{12}$, from the fluorescence of the L$_3$ shell for energies above $E_{L1}$, $f_{13}$ can be determined.
\FloatBarrier
\FloatBarrier

\section{Results}
\begin{figure}
  \centering
    \includegraphics[width=15cm]{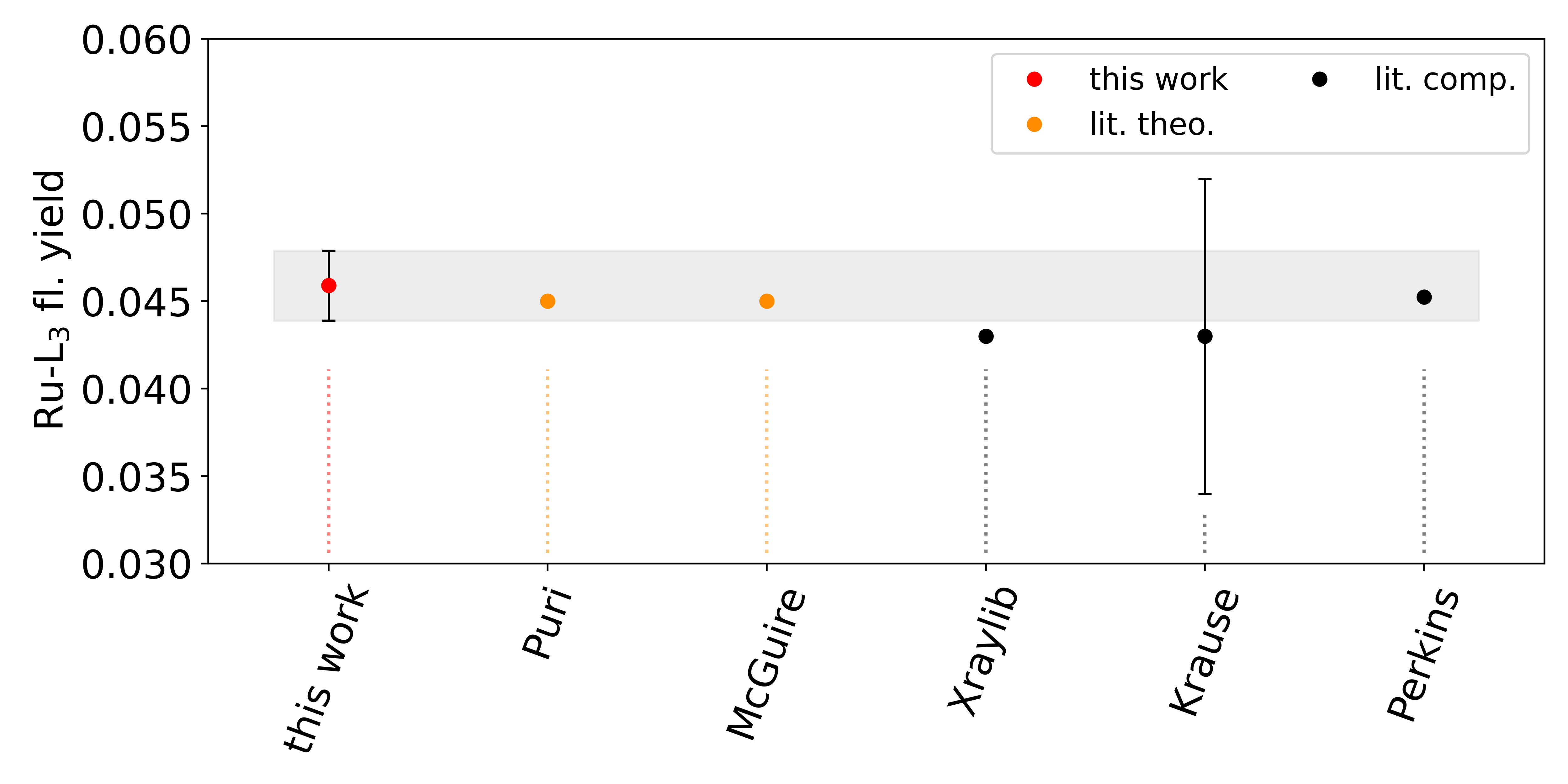}
    \includegraphics[width=15cm]{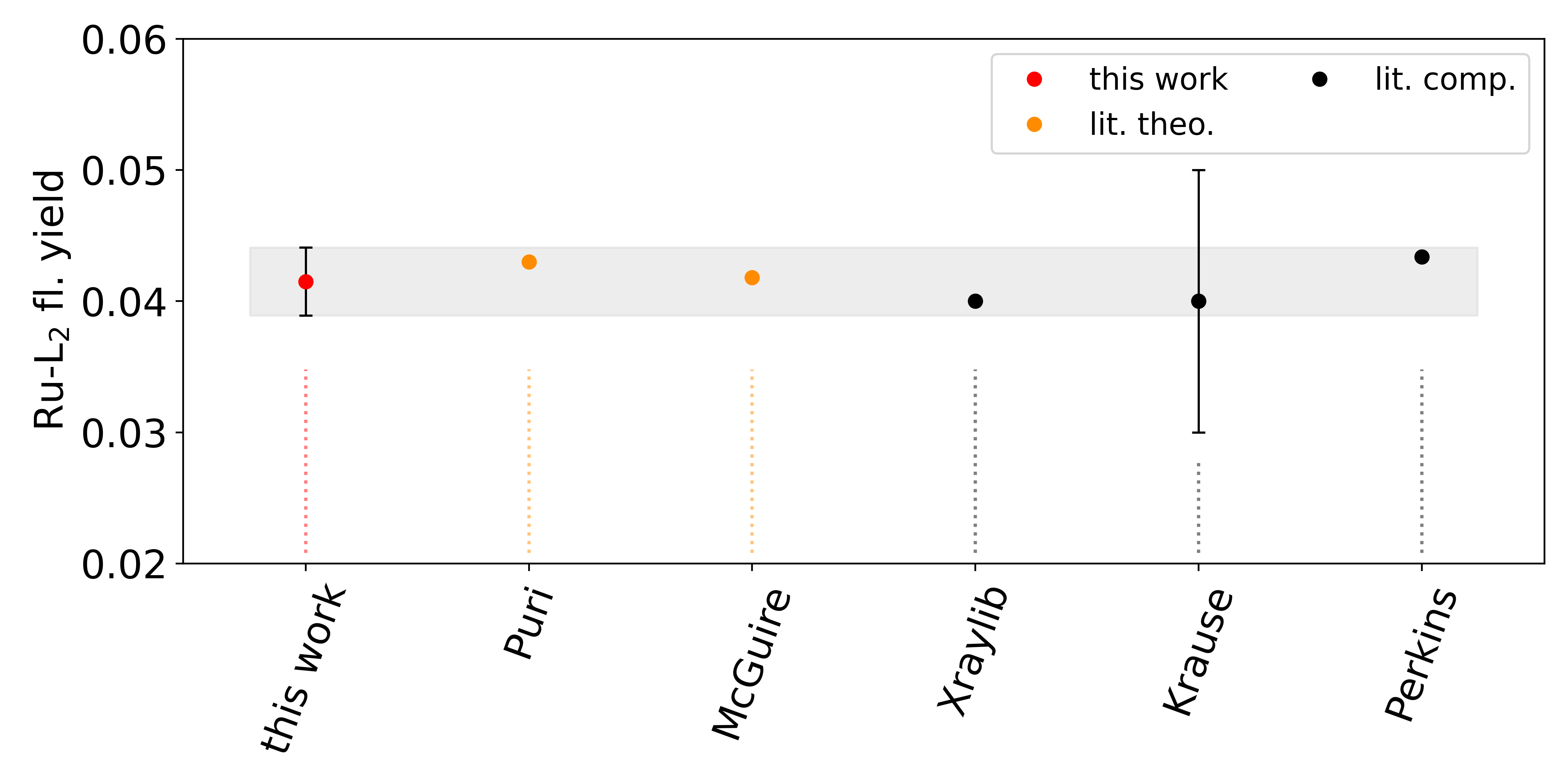}
    \includegraphics[width=15cm]{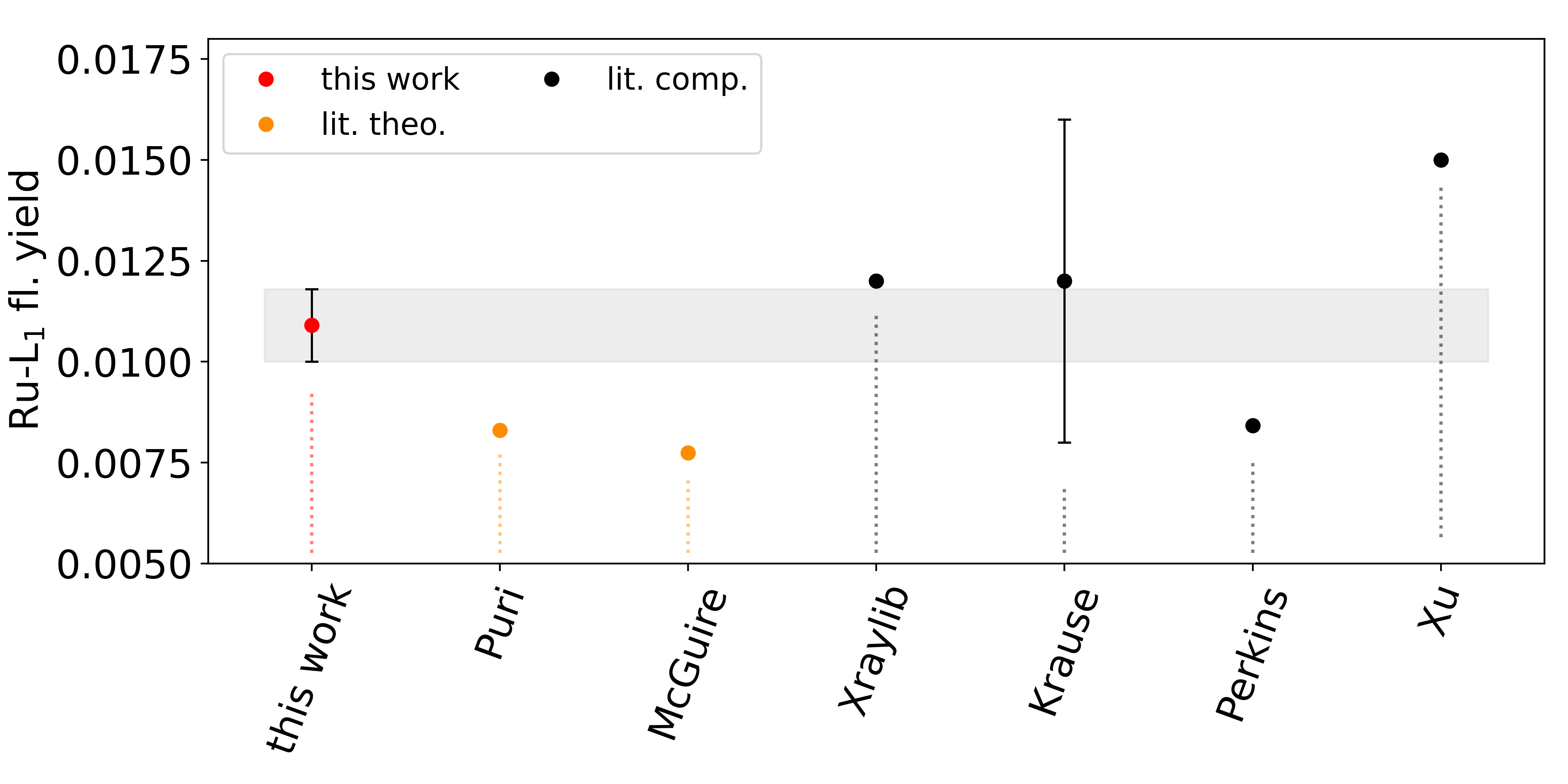}
  \caption{Comparison of the experimentally determined Ru-L-subshell fluorescence yields with values from the literature.}
  \label{fig:comparisonYields}
\end{figure}

\begin{figure}
  \centering
    \includegraphics[width=15cm]{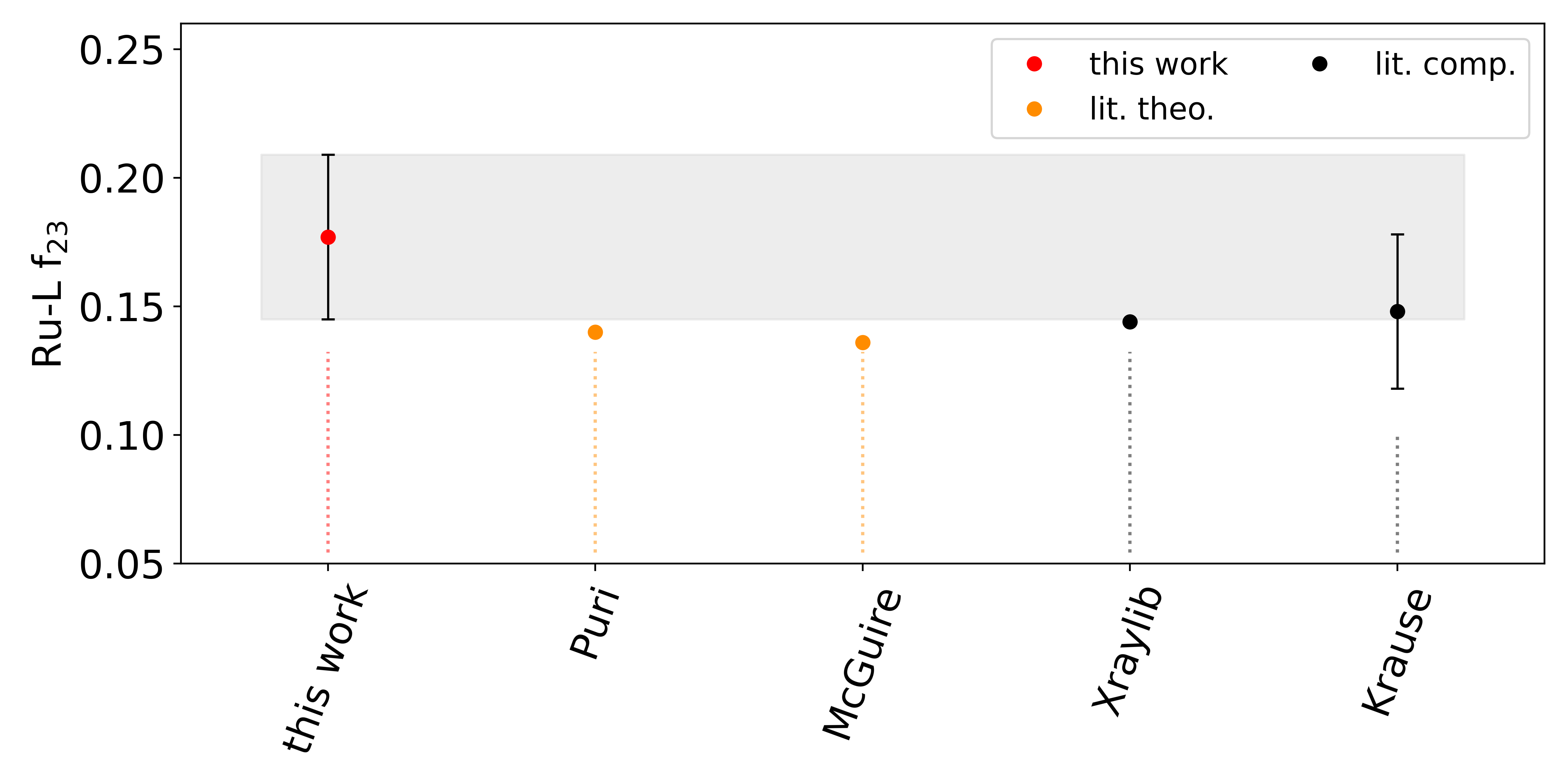}
  \caption{Comparison of the experimentally determined Coster-Kronig factor $f_{23}$ with values from the literature.}
  \label{fig:CKs}
\end{figure}

The determined fluorescence yields are $\omega_{L_3}=0.0459(20)$, $\omega_{L_2}=0.0415(26)$ and $\omega_{L_1}=0.0109(9)$. The resulting Coster-Kronig factors are $f_{23}=0.177(32)$, $f_{13}=0.528(91)$ and $f_{12}=0.173(73)$. These values are compared with values from the literature in table \ref{tab:table} and figure \ref{fig:comparisonYields} and \ref{fig:CKs}. The respective uncertainties were calculated via error propagation. The main contributions to the total uncertainty budget of the fluorescence yields were arising from the spectral deconvolution (\raisebox{-0.7ex}{\~{}}2 \%) and from the photoionization cross sections (\raisebox{-0.7ex}{\~{}}2 \%). The uncertainty budget is calculated by applying the reference-free XRF approach for the FP determination, discussed in more detail in \cite{Unterumsberger2018}.\\

The X-raylib \cite{T.Schoonjans2011} and Krause\cite{Krause1979} values of $\omega_{L3}$ and $\omega_{L1}$ are slightly outside of the error domain of the values determined in this work. The agreement with respect to the theoretically calculated data of Puri\cite{S.Puri1993} and McGuire\cite{McGuire} is better in the case of $\omega_{L3}$ but even worse for $\omega_{L1}$. The data of Perkins\cite{S.T.Perkins1991} as well as the data by Xu\cite{xu1987} behaves very similarly. For $\omega_{L2}$,  all available data from the literature agrees well with the result obtained here.
With respect to the Coster-Kronig factors, the tabulated data in X-raylib and the Krause compilation is in good agreement with our results. However, the results are on or slightly outside the boundary of our uncertainty budget for all three CK values. The data by McGuire and Puri is outside of our results considering their uncertainty budget.

\begin{table}
 \caption{Comparison of the experimentally determined Ru-L-subshell fluorescence yields and Coster-Kronig factors with the X-raylib database \cite[version 4.0.0]{T.Schoonjans2011} and other values (values from compilations (comp.) and theoretic values) from the literature.}
  \centering
  \begin{tabular}{c|c|c|c}
    \toprule
    \toprule
    & Ru $\omega_{L3}$     & Ru $\omega_{L2}$     & Ru $\omega_{L1}$ \\
    \midrule
    this work (XRF) & 0.0459(20) & 0.0415(26)  & 0.0109(9)     \\
    X-raylib \cite{T.Schoonjans2011} (comp.) & 0.043 & 0.040  & 0.012     \\
    Krause \cite{Krause1979} (comp.) & 0.043(9) & 0.040(10)  & 0.012(4)     \\
    Perkins et. al. \cite{S.T.Perkins1991} (comp.) & 0.045231 & 0.043368  &  0.0084138\\
    McGuire \cite{McGuire} (theory) & 0.0450 & 0.0418  &  0.00774\\
    Puri et. al. \cite{S.Puri1993} (theory) & 0.045 & 0.043  &  0.0083\\
    Xu et. al. \cite{xu1987} (comp.) & &  & 0.015\\
    \toprule
    \toprule
   &    Ru $f_{23}$     & Ru $f_{13}$     & Ru $f_{12}$ \\
    \midrule
    this work (XRF) &  0.177(32)     & 0.528(90)       & 0.173(73)  \\
    X-raylib \cite{T.Schoonjans2011} (comp.) &  0.144     & 0.61       & 0.10  \\
    Krause \cite{Krause1979} (comp.) & 0.148(30) & 0.61(7)  & 0.10(2)     \\
    McGuire \cite{McGuire} (theory) & 0.136 & 0.779  & 0.057 \\
    Puri et. al. \cite{S.Puri1993} (theory) & 0.140 & 0.766  &  0.057\\
   \bottomrule
  \end{tabular}
  \label{tab:table}
\end{table}

\section{Conclusion}
The Coster-Kronig factors and the fluorescence yields of ruthenium are determined experimentally by applying PTB's radiometrically calibrated instrumentation. The values determined are in reasonably good agreement with the values from the literature, although some literature values are slightly outside the uncertainty ranges of this work. The magnitude of the determined uncertainties of this work is much lower than the estimated uncertainties of Krause \cite{Krause1979} in the case of the fluorescence yield values. With respect to the Coster-Kronig factors, similar uncertainties were achieved here. In summary, this uncertainty reduction will positively influence the total uncertainties of fundamental parameter-based quantitative X-ray fluorescence experiments. \\
As stated already in previous works of our group \cite{Hoenicke2022, Unterumsberger2018, P.Hoenicke2016a, Ta-Paper}, the X-raylib database is also in the case of the Ru-L shell fundamental parameters a reliable reference.
\FloatBarrier
\section*{Conflict of interest}
There are no conflicts to declare. %https://www.rsc.org/journals-books-databases/author-and-reviewer-hub/authors-information/responsibilities/

\section*{Acknowledgments}
This project has received funding from the ECSEL Joint Undertaking (JU) IT2 under grant agreement No 875999. The JU receives support from the European Union’s Horizon 2020 research and innovation programme and the Netherlands, Belgium, Germany, France, Austria, Hungary, the United Kingdom, Romania and Israel.

%Bibliography
%\bibliographystyle{unsrt}  
\bibliographystyle{vancouver}
\bibliography{templateArxiv.bib}

\end{document}